The distribution of radio quiet active galactic nuclei in the star formation-stellar mass plane

David Garofalo[1] & George Mountrichas[2]

1. Department of Physics, Kennesaw State University, USA
2. Instituto de Fisica de Cantabria (CSIC – Universidad de Cantabria), Spain

Abstract

That active galactic nuclei (AGN) with jets can alternately enhance as well as suppress star formation rates, explains the location and slope of radio loud AGN on the star formation rate – stellar mass plane. Here, we explore 860 type 1 and 2 AGN at $z < 0.2$ from the ROSAT-2RXS survey in order to understand both different location and lower slopes for non-jetted AGN in the star formation rate – stellar mass plane. We describe the nature of these differences in terms of different degrees of black hole feedback, with relatively weak negative feedback from non-jetted AGN compared to both relatively strong positive and negative feedback from jetted AGN. The validity of these ideas brings us a step closer towards understanding the black hole scaling relations across space and time.

1. Introduction

While it is now recognized that orientation cannot be solely responsible for the differences between radio loud and radio quiet AGN (Urry & Padovani 1995), the conditions that determine whether a jet exists have not been fully identified. The radio loud/radio quiet dichotomy, in other words, remains a long-standing unresolved issue in high energy astrophysics. Some of the confusion comes from selection effects that appear to suggest a difference in the clustering of the two subgroups of AGN (Worpel et al 2013; Retana-Montenegro & Rottgering 2017), as well as a difference in the merger signatures of radio loud versus radio quiet AGN (Ivison et al 2012; Wylezalek et al 2013; Chiaberge et al 2015; Hilbert et al 2016; Noirot et al 2018; Zakamska et al 2019). There is also a sense that radio loud quasars have larger black hole mass compared to radio quiet ones (e.g. Laor 2000; Metcalf & Magliocchetti 2006).

In fact, for any cluster richness, radio quiet AGN are the majority (Garofalo, North, Belga, Waddell 2020), both populations are likely produced in mergers (Garofalo 2019; Garofalo, Webster, Bishop 2020), and radio quiet AGN form at the highest black hole mass (Oshlack et al 2002; Woo & Urry 2002; McLure & Jarvis 2004; Kelly et al 2008). What seems to be emerging is that radio loud AGN struggle to form unless the black hole is massive but that does not preclude most massive accreting black holes from being jetless or radio quiet (Garofalo, North, Belga, Waddell 2020). While mergers appear to trigger both subgroups of AGN, one difference seems to be that these subclasses of AGN live different timescales and therefore are observed having different strengths in their merger signatures (Garofalo 2019; Garofalo, Webster, Bishop 2020). Finally, the environmental preference of radio loud AGN for ellipticals compared to the ubiquitous distribution of radio quiet AGN in both spirals and ellipticals must be explained (Garofalo, North, Belga, Waddell 2020; Rusinek, Sikora, Koziel-Wierzbowska, Gupta 2020).

In this work, we explore the consequences of a picture for the radio loud/radio quiet dichotomy suggesting that radio quiet AGN produce different feedback. While radio loud AGN have the ability to both enhance and suppress star formation, the radio quiet ones can only suppress it, and, generally, to a lesser degree. We use observations from the ROSAT-2RXS survey on type 2 AGN at z < 0.2 to illustrate this difference on the star formation – stellar mass plane. In Section 2 we describe recent work on the behavior of jetted AGN and then relate that to the behavior of non-jetted AGN. In Section 3 we discuss the implications of the difference in the two populations and conclude.

2. The distribution of radio quiet AGN in the SFR-SM plane

We begin our exploration of radio quiet AGN by describing theoretical differences in their engines compared to radio loud AGN and then proceed to an understanding of how that difference explains their location on the star formation rate-stellar mass plane. The elements that characterize differences between radio loud and radio quiet AGN are state of accretion (radiatively efficient versus radiatively inefficient), black hole spin (high versus low), and direction of the disc angular momentum relative to that of the black hole (corotating versus counterrotating discs). The quantitative differences can be found in Garofalo, Evans & Sambruna (2010).

2.1 Radio quiet AGN

In Figure 1 we describe the time evolution of an initially moderately fast spinning black hole (spin $a$ = 0.9) surrounded by an accretion disc whose angular momentum is in the same direction as that of the black hole (lower panel). The cold gas that forms the radiatively efficient disc (shown in blue), comes from a merger. These are conditions that do not allow for the formation of a jet. Because no physics exists that can change the state of accretion, the black hole simply spins up as the angular momentum from the disc gas accretes onto the black hole. Eventually the black hole spins up to its maximum value and remains in that state until the cold gas reservoir runs out. Note how the inner edge of the disc moves closer to the black hole for higher spin as a result of the spin dependence of the stability of circular orbits. As the disc inner edge moves further in, the system taps deeper into the gravitational potential energy of the black hole, reprocesses that energy further out in the disc, and generates a stronger radiatively driven wind (Shakura & Sunyaev 1973; Kuncic & Bicknell 2007). This is shown in red. This wind is mildly capable of pushing the gas away and therefore reduces the star formation rate over time but it is not as effective in reducing the star formation rate as the direct heating of the ISM of an FRI jet as we will describe in Section 2.2. The bottom line is that radio quiet AGN that are triggered in a merger begin their life with some total stellar mass and specific star formation rate, and move mostly down in the SFR-SM plane due to disc wind - SFR suppression. When the reservoir of cold gas comes to an end, the AGN shuts down and the subsequent decrease in SFR is no longer associated with the AGN classification.

The simplicity of Figure 1 allows us to capture a characteristic or typical path for radio quiet AGN in the SFR-SM plane. To be specific, let us imagine the triggering of such an AGN in a more isolated environment that has a total initial stellar mass component corresponding to log (M$_*$/M$_\odot$) = 10.5 and an initial star formation rate of 0.315 solar masses/year. Over the course of $10^8$ years the SFR drops due to the disc wind so we can estimate an average SFR of 0.2 for that time period. This constitutes a relatively low drop in SFR compared to that associated with radio loud AGN. This gives an additional SM of 2 x $10^7$ M$_\odot$ and a total SM of 3.3 x $10^{10}$ M$_\odot$, which corresponds to log(SM) = 10.52. Hence, the increase on the horizontal axis is small compared to the decrease in the vertical axis. Regardless of whether we start at different SFR and SM values, the typical path for a radio quiet AGN that is subject to the evolution described in Figure 1, involves a mostly vertical drop in the SFR-SM plane. Again, this kind of behavior is weakly dependent on the initial location in the SFR-SM plane. The is due to a scale invariance that does not apply to radio loud AGN as we now proceed to describe.

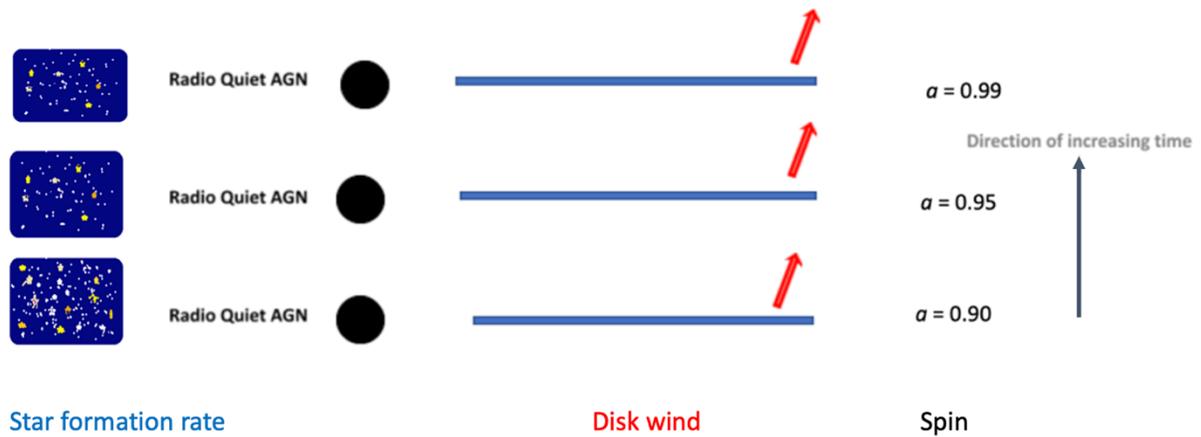

Figure 1: The evolution in time of an accretion disc that settles in corotation around a spinning black hole. Because the inner edge of the accretion disc moves closer to the black hole as the spin increases due to accretion, the disc wind becomes stronger with time which decreases the star formation rate shown on the left but otherwise has relatively weak impact on the system.

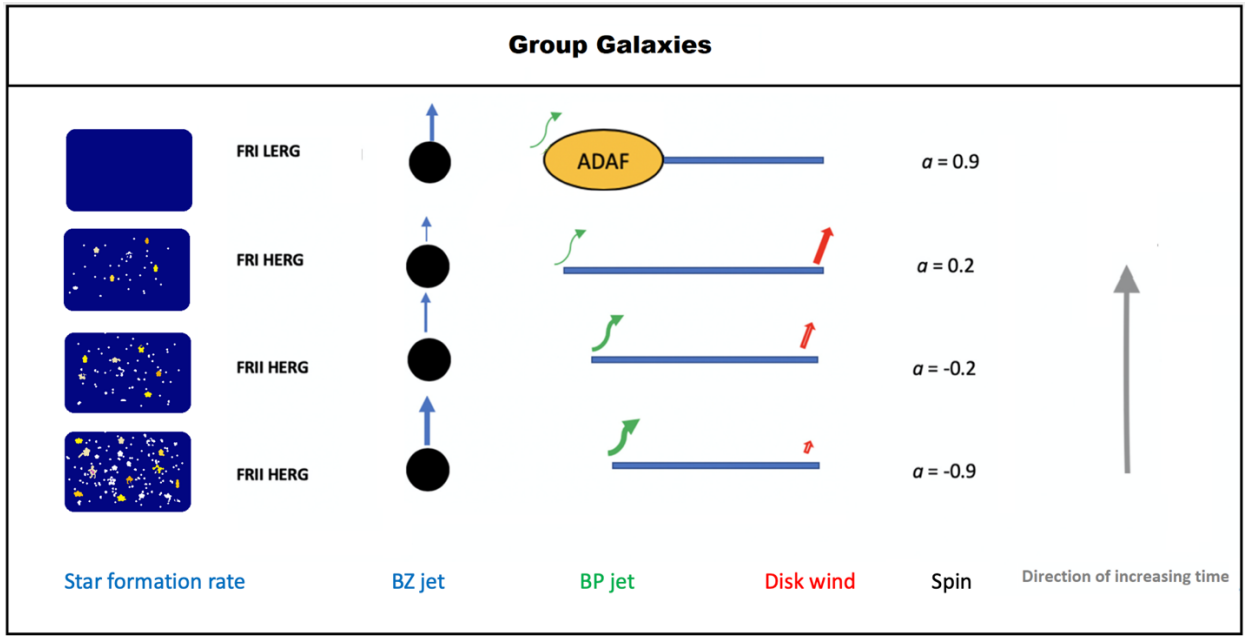

Figure 2: The evolution in time of jetted AGN in environments of intermediate density. The difference with respect to the lowest panel in Figure 1 is that the disc settles into counterrotation around the spinning black hole. As a result, the Blandford-Znajek (BZ - blue arrow) and Blandford-Payne (BP - green arrow) jets are maximized. ADAF refers to advection dominated accretion flow. Since the feedback is not as strong as in cluster environments, on average, the state of the disc evolves slowly. Left column shows the SFR suppression effect of the FRI jet some $10^8$ years after the AGN was triggered in a merger.

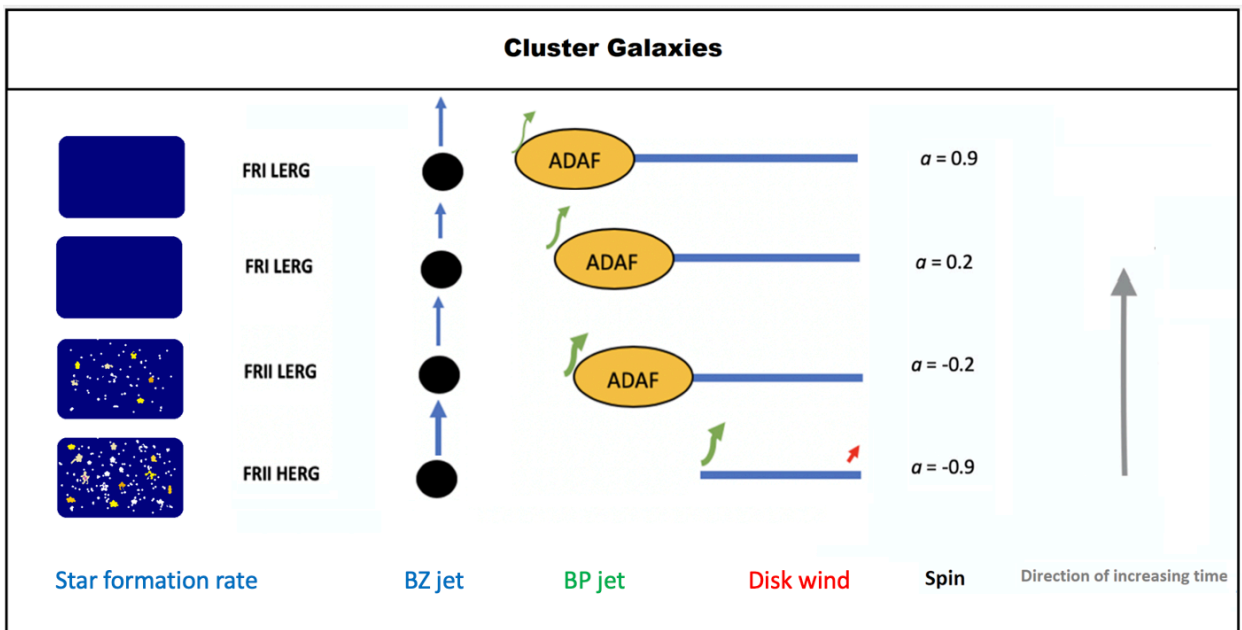

Figure 3: The evolution in time of jetted AGN in richest environments. The difference with respect to the lowest panel in Figure 1 is that the disc settles into counterrotation around the spinning black hole. The difference with respect to Figure 2 is that the black hole is more massive. As a result, the BZ and BP jets are maximized, and the feedback is stronger on average than in less dense environments. As a result, the disc evolves to an ADAF more quickly. Left column shows the effect on the SFR of the FRI jet.

2.2 Radio loud AGN

In Figures 2 and 3 we show the time evolution of post-merger accreting, high spinning black holes, whose accretion discs have settled in counterrotation about the black hole. This leads to fundamental differences with respect to Figure 1 systems which we describe qualitatively since our focus here is on the radio quiet AGN. For a more in-depth, quantitative analysis, of the radio loud subgroup of AGN see Singh et al (2021). The difference between the lower panels of Figures 2 and 3, and the lower panel of Figure 1, is counterrotation versus corotation following a gas rich merger. In counterrotation, the inner edge of the disc lives further away from the black hole. This larger size in the gap region between the disc and the black hole maximizes the Blandford-Znajek and Blandford-Payne jets, shown in blue and green, respectively. Maximizing these two jet mechanisms in tandem leads to a collimated FRII jet. Because the discs are thin and radiatively efficient (in blue), the system is in high excitation and thus labeled HERG for high excitation radio galaxy. Because the black holes tend to be larger for denser environments, the feedback in these environments is more effective, and on average the accretion disc changes state earlier. This is shown in the second to bottom panel of Figure 3, with an advection dominated accretion flow (ADAF) replacing the thin disc. Jetted AGN experience two distinct phases in the coupling of jets to star formation. In the initial FRII phase, the jet enhances star formation, while in the FRI phase, it suppresses the star formation rate. The reason for the suppression of star formation by the FRI jet is due to the tilt in the jet that results from the absence of the Bardeen-Petterson effect as the black hole transitions through zero spin (Garofalo, Joshi et al 2019). This is referred to as the Roy Conjecture (from Namrata Roy – Garofalo et al in preparation). The left column indicates the star formation rate. But since richer environments evolve to ADAF accretion states rapidly, on average, their evolution is slowed down compared to less dense environments as a result of the fact that ADAF accretion involves accretion rates that are at least two order of magnitude lower than for thin discs. As a result, when a typical accreting black hole spins down to zero spin in a rich environment via ADAF accretion, a black hole formed at the same time that does not transition to an ADAF early, will have already spun its black hole up into the corotating phase. This is shown in Figure 4, with A indicating the location in the SFR-SM plane for an original counterrotating black hole in a less dense environment formed at the same time as a black hole that is at location B in a richer environment. Location B is the transition between the counterrotating phase and the corotating phase (at zero black hole spin).

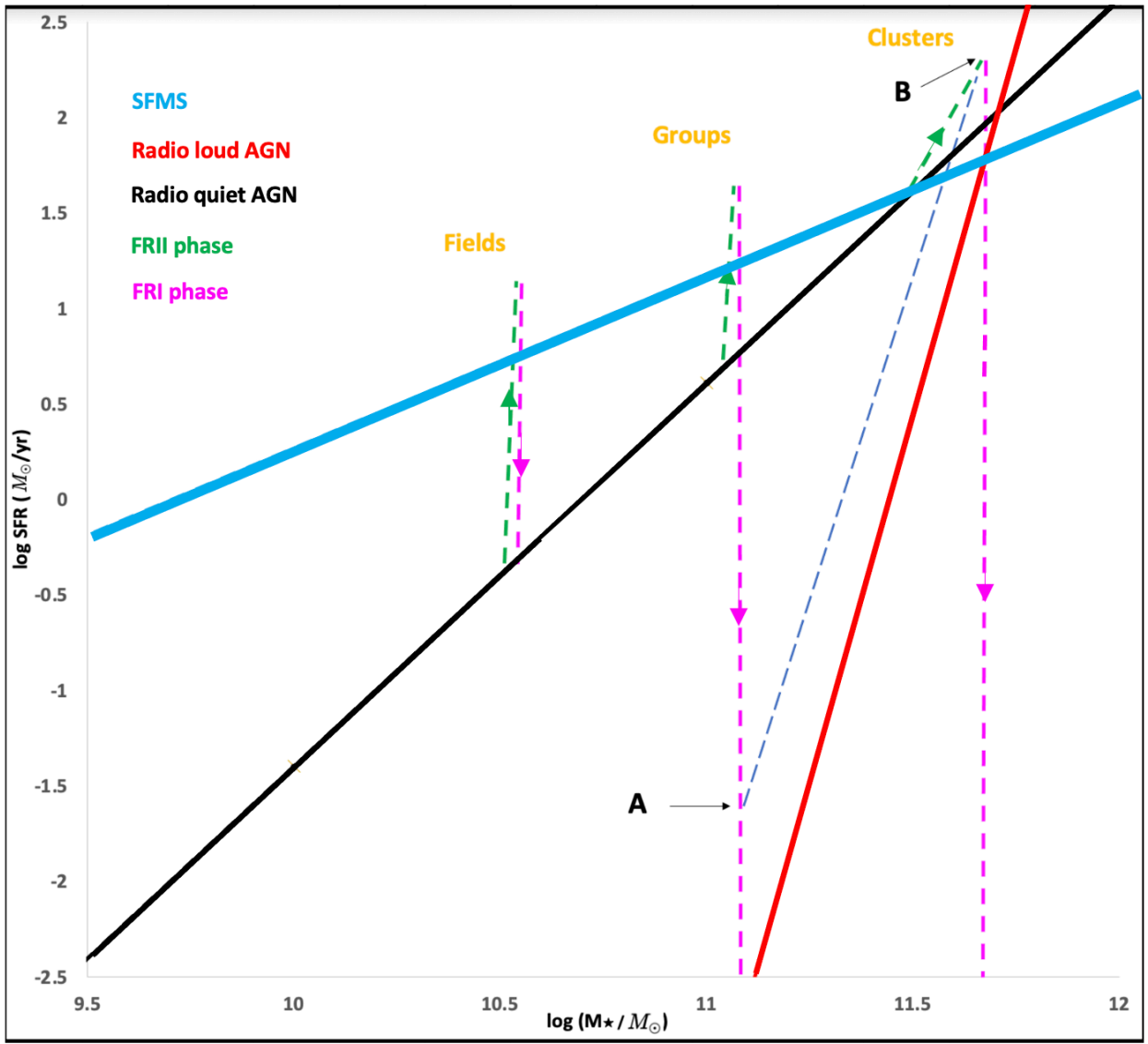

Figure 4: The paths of radio loud AGN on the SFR-SM plane from Singh et al (2021), and a characteristic late time location for an object triggered in a group environment (point A) as compared at the same time to the location of a characteristic object triggered in a cluster environment (point B). The two radio loud AGN triggered at the same time in group and cluster environments evolve at different rates with the former tending to evolve quickly in the SFR-SM plane, while the latter more slowly. As a result, the less isolated AGN will tend to be found when it is deeper in its evolution at point A compared to point B for the AGN in richest environment. This difference in timescales for evolution explains the observed slope (blue dashed line). The radio loud, and radio quiet fits come from Comerford et al 2020 while the SFMS is from Elbaz et al 2007.

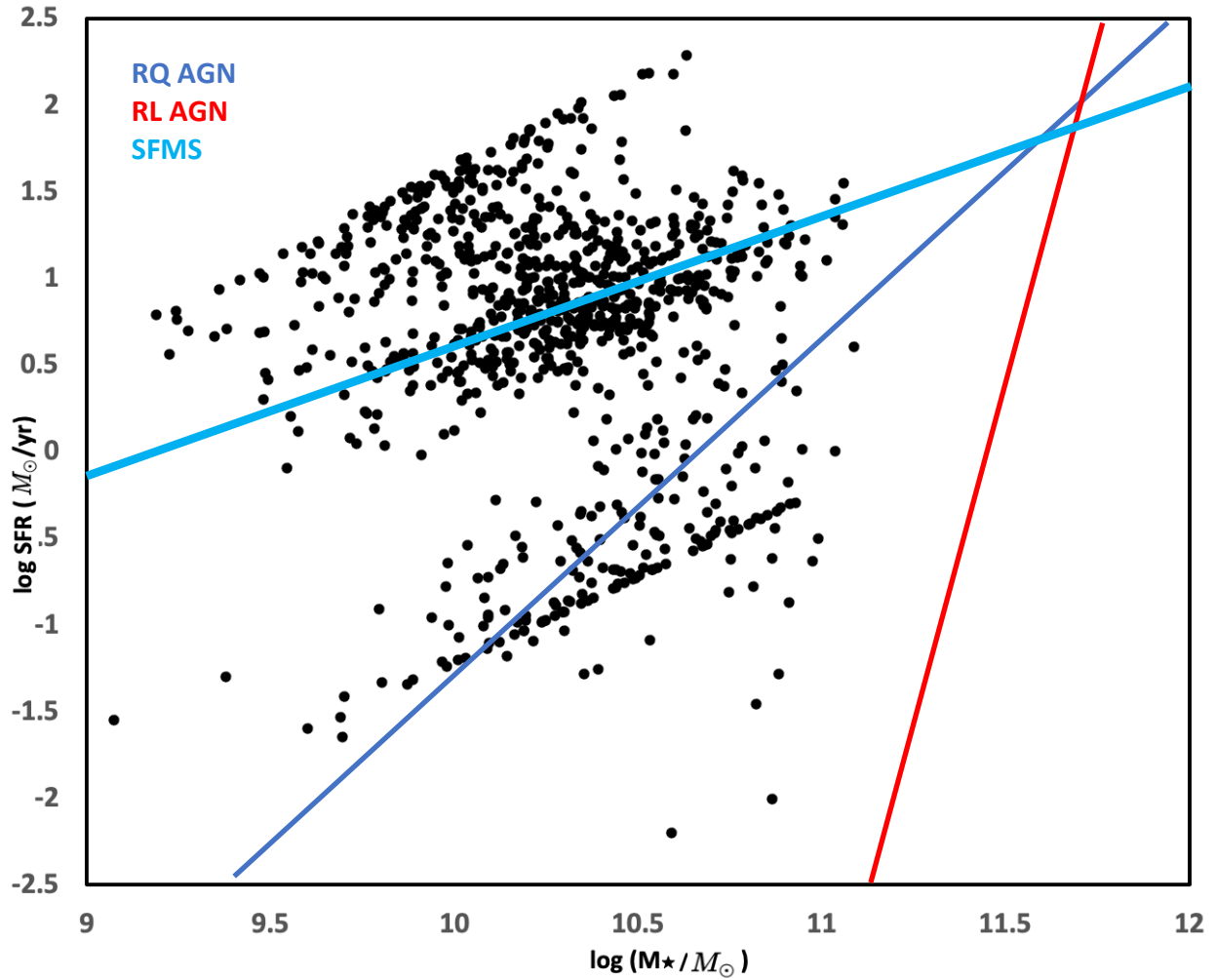

Figure 5: The 860 type 1 and 2 AGN at z < 0.2 from the ROSAT-2RXS survey in the SFR-SM plane in black. For comparison, we added the fit of the radio quiet AGN (in blue), the fit for the radio loud AGN (in red) of Comerford et al 2020 and the SFMS (Elbaz et al 2007 in light blue) showing different location and different slope for the two AGN subclasses.

The key point to take from our synthesis of the radio loud AGN paths in the SFR-SM plane is that more radio loud AGN will be found in the middle of the diagram closer to point A compared to locations on the diagram with the same SFR but larger SM. Equivalently, if we start at point B and move horizontally to smaller SM values, we expect to find fewer RL AGN compared to near point B. This is due to rapid and slow evolution through counterrotation and into corotation in less dense environments and therefore to different likelihood of finding the object at specific SFR-SM values. Hence, the slope of the radio loud AGN will tend to be steep as it connects objects near A and near B.

3. Discussion and conclusion

In Figure 5 we plot 860 type 1 and 2 AGN at z < 0.2 from the ROSAT-2RXS survey in black (which were selected out of a sample of 1021 – see section 2 of Koutoulidis et al 2021), and the fit of the radio quiet AGN and radio loud AGN from Comerford et al (2020) in order to expand the comparison between model and the radio loud AGN based on the Comerford et al (2020) work to include a radio quiet AGN sample. We use the host galaxy properties (SFR and SM) calculated in Koutoulidis et al 2022. For their estimation, spectral energy distributions (SEDs) have been constructed using optical (SDSS), near-IR (Skrutsie et al 2006) and mid-IR (Wright et al 2010) photometry and have been fit with the X-CIGALE algorithm (Boquien 2019, Yang 2020). Although starbursts are rare at low redshift, there were a few sources at higher SFR, but they had unreliable SED fits and were excluded by the quality selection criteria. The small concentration of sources at low SFR across the plot is difficult to completely avoid and is caused by the parameter space of the parameters of the star formation history module. A denser grid in e.g. the stellar age parameter at high(er) values would render the line blurry. Nonetheless, it does not affect our conclusions. The templates and the parameter grid utilized, is described in detail in Section 3.1 of Koutoulidis et al 2022.

In this work, we use only sources with reliable SFR and SM calculations, applying the criteria described in Section 3.2 of Koutoulidis et al (2022). Specifically, we exclude sources whose fit has reduced $\chi^2$ greater than 5 (Masoura et al 2018; Mountrichas et al 2019, Buat et al 2021). We also make use of the ability of X-CIGALE to calculate two values for each estimated parameter. One is calculated from the best-fit model and one that weighs all models allowed by the parametric grid, with the best-fit model having the heaviest weight (Boquien et al 2019). This weight is based on the likelihood, $\exp(-\chi^2/2)$, associated with each model. A large difference between these two values, indicates that the fitting process did not result in a reliable estimation for this parameter (Mountrichas et al 2021b; Mountrichas et al 2021c; Buat et al 2021). Thus, we only include in our analysis sources with

$$1/5 \leq M_{*,best}/M_{*,bayes} \leq 5 \quad \text{and} \quad 1/5 \leq SFR_{best}/SFR_{bayes} \leq 5$$

where $M_{*,best}$, $SFR_{best}$ and $M_{*,bayes}$, $SFR_{bayes}$ are the best and Bayesian fit values of SM and SFR, respectively. 860 X-ray AGN meet these requirements.

We also add the fit of the radio quiet AGN and radio loud AGN from Comerford et al (2020). Clearly, the slope of our data is smallest. In addition to the differences in slope, the red line is shifted to the right and downward, indicating that the radio loud AGN tend to occupy a region rightward as well as downward of the radio quiet ones. We can understand this by recognizing that the radio loud AGN that experience the pink FRI, star formation suppression phase, are black holes accreting at increasingly low rates (Figure 3, top panel and Figure 4 pink phase of cluster RL AGN). Because of the low accretion rates, such objects remain within the radio loud AGN classification as the SFR values drop to extremes. Radio quiet AGN, on the other hand, do not experience ADAF accretion states (otherwise they would exit the radio quiet AGN classification – Figure 1), so they accrete at near Eddington values. As a result, they consume their fuel much faster than the FRI radio galaxies and therefore exit the AGN classification altogether before SFR values can become very low.

The simple time evolution described in Figures 1, 2, and 3, allows one to appreciate the upwards and downwards paths, with slight rightward motion, for radio loud AGN in the SFR-SM plane. When the FRII jet feedback is relatively less effective, the accretion state evolves slowly into an ADAF, and this allows the system to reach its pink FRI phase more rapidly, where such systems remain as long as the accretion fuel is available. Hence, it is in these intermediate density environments that we expect the likelihood of finding radio loud AGN to increase. In richer environments, on the other hand, the time evolution during the FRII phase is slowed down by two orders of magnitude or more, as a result of the more effective FRII jet feedback, and the resulting rapid evolution from a thin disc to a radiatively inefficient ADAF one. As a result, the average system in the densest cluster environments will have accreting black holes that linger longer near the transition from FRII to FRI phases. This difference in the characteristic evolution timescales produces the steep slope for radio loud AGN in the SFR-SM plane. The absence of jets for radio quiet AGN makes their characteristic evolution scale-invariant, in the sense that massive and less massive black holes simply lower the SFR in their host galaxies in a way that is proportional to their disc winds (Figure 1). The absence of ADAF accretion forces such systems on average to consume their fuel orders of magnitude more rapidly compared to radio loud AGN in dense environments. The combination of shorter lifetimes and weaker feedback in non-jetted AGN, allows for SFR to drop less compared to FRI phases in jetted AGN. Of course, we are not predicting the nature of individual objects, but about the overall character of these 860 type 1 and 2 AGN. Given how different the feedback is prescribed to be in radio loud versus radio quiet AGN, it is conceivable that we will understand how radio loud AGN will display black hole scaling relations that are quite different from those of radio quiet AGN. And, these scaling relations should be dynamic and therefore redshift dependent. And the SKA should allow us to explore the crucial low power transition radio AGN at higher redshift such as X-shaped radio galaxies and FR0 radio galaxies needed to understand time evolution.

Acknowledgments

We thank the anonymous referee for a thorough and detailed review. GM acknowledges support by the Agencia Estatal de Investigación, Unidad de Excelencia María de Maeztu, ref. MDM-2017-0765.